\documentclass[preprint,longabstract]{aastex}
\usepackage{natbib}
\usepackage{multirow}
\usepackage{color,graphicx}
\usepackage{hyperref}

\newcommand{\sur} {\mathrm{(\emph{u}}-\mathrm{\emph{r})}}
\newcommand{\dinbas}{DynBaS}

\newcommand{\dinbasdos}{DynBaS2D}
\newcommand{\dinbastres}{DynBaS3D}
\newcommand{\dinbasnd}{DynBaSND}

\newcommand{\gaspex}{GASPEX}
\newcommand{\tgaspex}{TGASPEX}
\newcommand{\ssag}{SSAG}
\newcommand{\starlight}{{\tt STARLIGHT}}
\newcommand{\ch} {$\chi^2$}
\shorttitle{Galaxy properties from SED fitting}
\shortauthors{Magris et al.}

\begin{document}

\title{On the recovery of galaxy properties from SED fitting solutions}

\author{Gladis~Magris C.}
\affil{Centro de Investigaciones de Astronom\'{\i}a, AP 264, M\'erida 5101-A, Venezuela}
\email{magris@cida.gob.ve}

\author{Juan~Mateu P.\altaffilmark{1}}
\affil{Departamento de F\'isica, FACYT, Universidad de Carabobo, Valencia, Venezuela}
\email{mateu@uc.edu.ve}

\author{Cecilia~Mateu\altaffilmark{2}}
\affil{Instituto de Astronom\'ia, UNAM, Ensenada, C.P. 22860, Baja California,  M\'exico}
\email{cmateu@astrosen.unam.mx}

\author{Gustavo~Bruzual A.}
\affil{Centro de Radioastronom\'{\i}a y Astrof\'{\i}sica, CRyA, UNAM, Campus Morelia, A.P. 3-72, C.P. 58089, Morelia, Michoac\'an, M\'exico}
\email{g.bruzual@crya.unam.mx}

\author{Ivan~Cabrera-Ziri}
\affil{Astrophysics Research Institute, LJMU, 146 Brownlow Hill, Liverpool L3 5RF, UK}
\email{I.CabreraZiriCastro@2013.ljmu.ac.uk}

\and

\author{Alfredo Mej\'{i}a-Narv\'aez\altaffilmark{1}}
\affil{Centro de Investigaciones de Astronom\'{\i}a, AP 264, M\'erida 5101-A, Venezuela}
\email{mejia@cida.gob.ve}

\altaffiltext{1}{Postgrado de F\'{i}sica Fundamental, Universidad de Los Andes, M\'erida, Venezuela}
\altaffiltext{2}{Centro de Investigaciones de Astronom\'{\i}a, AP 264, M\'erida 5101-A, Venezuela}

\label{firstpage}

\begin{abstract}

We explore the ability of four different inverse population synthesis codes to recover the physical properties of galaxies from their
spectra by SED fitting.
Three codes, \dinbas, \tgaspex,~and \gaspex, have been implemented by the authors and are described in detail in the paper.
 \starlight, the fourth code, is publicly available.
\dinbas\ selects dynamically a different spectral basis to expand the spectrum of each target galaxy; \tgaspex\ uses an unconstrained age basis, whereas \gaspex ~and \starlight\  use for all fits a fixed spectral basis selected a priori by the code developers.
Variable and unconstrained basis reflect the peculiarities of the fitted spectrum and allow for simple and robust solutions to the problem of extracting galaxy parameters from spectral fits.
We assemble a Synthetic Spectral Atlas of Galaxies (\ssag)\footnote{Available at \url{http://www.astro.ljmu.ac.uk/~asticabr/SSAG.html}.}, comprising 100,000 galaxy spectra corresponding to an equal number of star
formation histories based on the recipe of \citet{chen12}. 
We select a subset of 120 galaxies from \ssag~with a colour distribution similar to that of local galaxies in the seventh data release (DR7) of the Sloan Digital Sky Survey (SDSS) and produce 30 random noise realisations for each of these spectra.
For each spectrum we recover the mass, mean age, metallicity, internal dust extinction, and velocity dispersion characterizing  the dominant stellar population in the problem galaxy.
All methods produce almost perfect fits to the target spectrum, but the recovered physical parameters can differ significantly.
Our tests provide a quantitative measure of the accuracy and precision with which these parameters are recovered by each method. 
From a statistical point of view all methods yield similar precisions, whereas \dinbas\ produces solutions with minimal systematic biases in the distributions of residuals for all of these parameters. We caution the reader that the results obtained in our consistency tests represent lower limits to the uncertainties in parameter determination.
Our tests compare theoretical galaxy spectra built from the same synthesis models used in the fits. 
Using different synthesis models and the lack of particular stellar types in the synthesis models but present in real galaxies will increase these errors considerably.
Additional sources of error expected to be present in real galaxy spectra are not easy to emulate, and again will result in larger errors.

\end{abstract}


\section{Introduction and Motivation}\label{s:intro}

\begin{figure}
\plotone{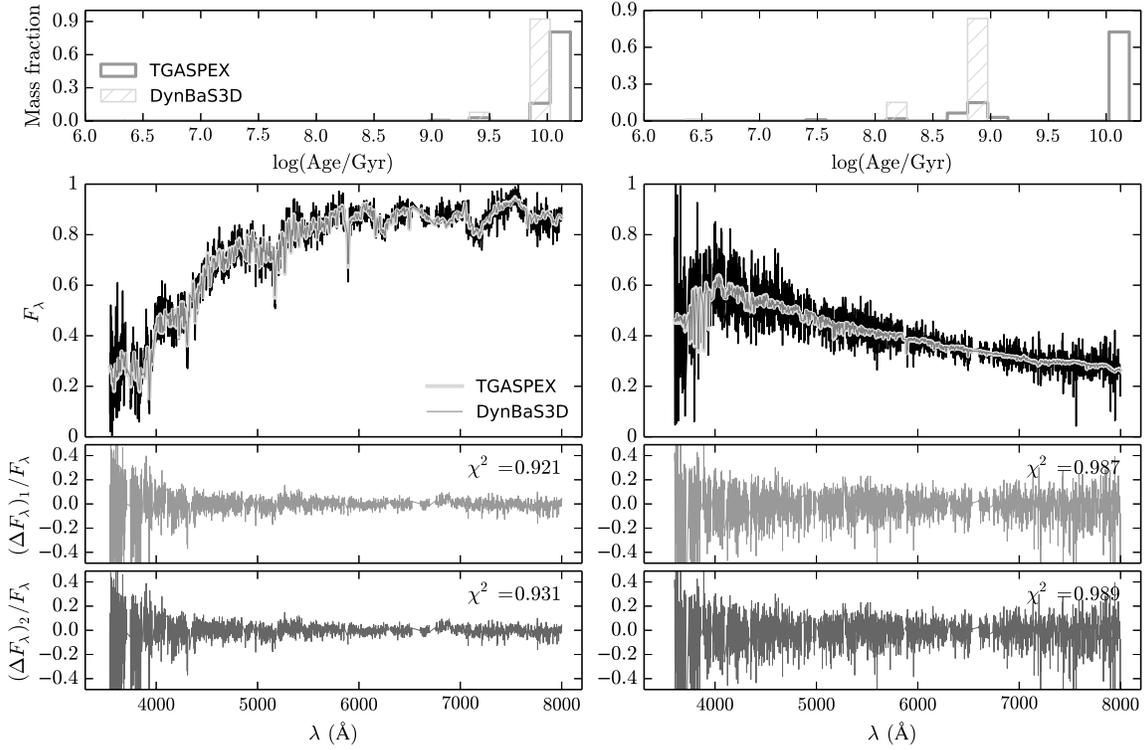}
\caption{SED fits obtained with \dinbastres\ and \tgaspex\ for two SDSS galaxies: an early type (\emph{left panels}) and a late type galaxy (\emph{right panels}). 
The main panel shows the spectral fits for both algorithms.
The two bottom panels show the fractional residuals $\Delta F_\lambda/F_\lambda=(F_\lambda-F_\lambda^{fit})/F_\lambda$ for 
\dinbastres\ and \tgaspex, respectively. 
The reduced $\chi^2$ for the fit is shown in the top right corner of the bottom panels.
The top panels show the fraction of the galaxy mass in each population component as a function of (log) age. See text for details.
}
\label{motiv}
\end{figure}

The stellar content of unresolved galaxies can be studied only through integrated photometry and/or spectroscopy.
The advent of large surveys centred at several rest-frame spectral ranges has encouraged the development of
tools to characterize the stellar populations dominating the light in these galaxies.
The literature of the last ten years is rich in papers on {\it inverse population synthesis}, which use spectral energy distribution (SED)-fitting
and parameter extraction algorithms as a natural application of population synthesis to derive the star formation history (SFH) of distant and nearby galaxies.
See papers by \citet{heavens00}, \citet{tojeiro07}, \citet{ocvirck06}, \citet{cidfernandes05},  \citet{koleva09} for details on different methods.

SED fitting refers to the use of different algorithms to select from a set of spectral evolution models, e.g. \citet[hereafter BC03]{bc03},
a single age spectrum, or a linear combination of spectra of different age, to construct the synthetic spectrum $F^{syn}_{\lambda}$ that minimizes the goodness-of-fit, e.g.,
\begin{equation}
\chi^2=\sum_i \frac  {(F^{obs}_{\lambda_i}-F^{syn}_{\lambda_i})^2} {\sigma_{\lambda_i}^2}.
\label{chi0}
\end{equation}
Here ${\sigma_{\lambda}^2}$ is the standard deviation of the gaussian distribution characterising the error in the flux of the target spectrum $F^{obs}_\lambda$
at wavelength ${\lambda}$.
Since the quality of the spectral fit, measured by \ch~ in Eq.~(\ref{chi0}), depends on the adequacy of the spectral evolution model to describe the problem spectrum,
most SED-fitting algorithms will produce equally good fits to the target spectrum.
However, flexible SED-fitting codes, that use a relatively large and variable number of individual spectra to build $F^{syn}_{\lambda}$ will, in general, produce
better fits to $F^{obs}_\lambda$ than the more rigid codes.

The ultimate goal of inverse modelling is not the spectral fit itself, but the interpretation of the components of the synthetic spectrum which minimize \ch\ in terms of the SFH of the problem galaxy, i.e. the run of star formation with time over the galaxy lifetime.
The derived spectral components can then be used together with the evolutionary synthesis models to recover basic physical parameters of the stellar population in this galaxy, e.g. stellar mass, mass-weighted-age, flux-weighted-age, average metallicity, mass-to-light ratio, dust content.
Parameter extraction is model dependent, and different algorithms can recover different values for these parameters from equally good spectral fits.

As motivation for our study, we show in Figure \ref{motiv} the results of fitting the SED of two SDSS DR7 galaxies \citep{abazajian09} using two different algorithms: \dinbastres\ and \tgaspex\ (these are introduced in \S \ref{threealgs}).
The fitted SEDs are indistinguishable in the scale of these plots (main panel).
This is illustrated also by the residual plots (two bottom panels) and evidenced by the minute differences in the values of reduced  \ch.
However, the stellar mass derived for the early-type galaxy (left panels) using
\tgaspex\ 
is 30\% larger than the mass obtained with
\dinbastres.
For the late-type galaxy in the right hand side panels the effect is even more striking.
The difference in reduced \ch\ is only 0.002, but the derived mass differs by 110\%. 
In the top panels of Figure \ref{motiv} we show for both solutions the fraction of the galaxy mass in each of the selected population components vs. log age. This function can be interpreted as the SFH, as will be discussed in a forthcoming paper.
For the early type galaxy both algorithms give approximately the same mass distribution, resulting in a {\it small} difference in the derived total mass.
For the bluer galaxy, \dinbastres\ gives a solution with a younger component than \tgaspex, and the inferred galaxy mass differ by a factor of two.
Several degeneracies present in spectral evolution models contribute to this ambiguity.
The discrepant results do not depend in a obvious manner on the signal-to-noise-ratio (SNR) of the target spectrum as we discuss in \S 3.

One needs then a quantitative measurement of the uncertainty in the derived galaxy physical parameters that takes into account these model limitations,
whose origin has been discussed in the papers cited above, as well as in the review paper by \citet{walcher_etal_11}.
In summary, star formation and evolution are complex processes, which overlap in galaxies over time scales ranging from a few million to billions of years.
These processes are highly simplified in stellar population synthesis models.
Additionally, by the nature of stellar evolution, at old ages the evolution is slow in almost all observable spectral features.
As a consequence, the footprints of evolution are blurred in integrated observational properties,
especially when multiple generations of stars are involved.
Despite these limitations, population synthesis modelling and its complementary SED-fitting and parameter extraction algorithms,
have been improved remarkably, in parallel with the quality of the large amount of observational material available today.

Inverse population synthesis algorithms can be classified as either parametric or non-parametric.
In the parametric case the galaxy SFH is assumed to follow an analytic expression that depends on one or more parameters \citep[see e.g.][]{dacunha08,acquaviva11,lee10,maraston10,guaita11}.
It is common practice to assume that galaxies begin to form stars at time $t_{form}$ according to an exponentially declining or increasing star formation rate (SFR) with $e$-folding time $\tau$ (which we will refer to as $\tau$-models). The goal in this case is to determine from the spectral fit the best values of  the parameters $t_{form}$ and $\tau$.
In the non-parametric case no functional form is assumed for the SFR. The galaxy SFH is then recovered from the stellar mass formed at each age bin in a previously defined grid of time steps, e.g. MOPED, VESPA, STECMAP, \starlight, ULySS \citep{heavens00, tojeiro07, ocvirck06, cidfernandes05, koleva09}.

In general, the SFH of a galaxy cannot be inferred with the same time resolution available in population synthesis models.
Testing their algorithm on spectra similar to the SDSS galaxy spectra, \citet{cidfernandes05}
concluded that a robust description of the SFH is only achievable in terms of three populations: young, intermediate and old. A discussion on this issue can also be found in \citet{cidfernandes14}.
Similarly, \citet{tojeiro07}, using their code VESPA, report that it is possible to recover a maximum of between 2 and 5
independent stellar population components from a pre-selected set of 16 age bins.
Both these algorithms use a fixed-age-basis of model galaxy spectra to decompose the target spectrum in Eq.~(\ref{chi0}).

Our goal in this paper is to characterize the uncertainties in the physical parameters describing the SFH of a galaxy derived
from non-parametric SED-fitting algorithms, and to provide the user with the typical error and bias associated to these parameters.
The paper is structured as follows.
In \S 2 we present the details of \gaspex, \tgaspex\, and \dinbas, our non-parametric SED fitting algorithms.
In \S 3 we contrast the performance of these methods on target spectra from a library of theoretical spectra (\ssag) based on the SFH recipes of \citet{chen12} to simulate SDSS spectra of galaxies. The details of this physically motivated spectral library are described in Appendix B.  We compare our results with those obtained with the
well known public code \starlight~\citep{cidfernandes05, cidfernandes14}.
In \S 4 we discuss the advantage of using flexible over fixed-age-basis algorithms and present the conclusions of our study.

\section{The Stellar Population Content of Galaxies}

Following BC03, the SED at time {\it t} of a
composite stellar population in which stars are formed with a constant initial mass function (IMF), can be synthesized from

\begin{equation}
 F_\lambda(t)\ =\ \int _0^t \Psi(t-t')\ f_\lambda[t',\zeta(t-t')]\ dt' ,
 \label{eq:cspi}
 \end{equation}

\noindent where $\Psi(t)$ and $\zeta(t)$ are the SFR and the metal-enrichment law, respectively.
$f_\lambda(t,Z)$ is the SED at age {\it t} of a simple stellar population (SSP) of metallicity $Z$ for the indicated IMF.
Several sets of spectra for SSPs are available in the literature, e.g. BC03 models.
These models provide a set of evolving spectra computed at a discrete number of time steps.
Thus, it is more appropriate to write Eq.~(\ref{eq:cspi}) as a linear combination of spectra:

\begin{equation}
F_\lambda(t)\ =\ \sum _{i,j}\ a_{i,j}\ f_\lambda(t_i,Z_{j}),
\label{flamsum}
\end{equation}

\noindent where the double sum is formally performed over the $N_t$ time steps $t_i$ available in the SSP model for metallicity $Z_j$ of a set of $K_Z$ metallicities, i.e. $j=1,...,K_Z$. We can think of Eq.~(\ref{flamsum}) as a sum in vector space.
The coefficient $a_{i,j}$ represents the weight assigned to the basis vector $f_\lambda(t_i,Z_j)$ in this space.
The SFR at time $(t-t_i)$ can then be derived directly from $\Psi(t-t_i) = a_i / \Delta t_i$, where ${a_i}=\sum_j a_{i,j}$, and
 $\Delta t_i= (t_{i+1}-t_{i-1})/2$.

\subsection{SED fitting}

Spectral fitting techniques seek to determine the SFH of a galaxy, i.e. $\Psi(t)$ and $\zeta(t)$, by minimizing the difference between the target
spectrum $F^{obs}_\lambda$ and a model spectrum $F^{mod}_{\lambda}$.
For these fits we can either use models that follow a parametric SFR, e.g. $\tau$-models with $\Psi(t) \propto \exp^{-t/\tau}$,
or make no a priori assumption about the distribution of ${a_{i,j}}$ in Eq.~(\ref{flamsum}).
In the latter case, the SFH can be estimated by computing the non-negative values of the coefficients ${a_{i,j}}$ that minimize the merit function
\begin{equation}
\chi^2=\sum_l \frac{[F^{obs}_{\lambda_l}-\sum _{i,j} a_{i,j} f_{\lambda_l}(t_i,Z_{j})]^2}{\sigma{_l^2}},
\label{chi2}
\end{equation}
used to measure the goodness-of-fit. The sum in Eq.~(\ref{chi2}) extends over all the wavelength points $\lambda_l$ in common in both spectra, and 
$\sigma_l$ is the standard deviation characterizing the error in the flux $F^{obs}_{\lambda_l}$, assumed to follow a gaussian distribution.

In this paper we follow the non-parametric approach, in which the SFH is not expressed as a function of one or more parameters.
For the $f_\lambda(t_i,Z_j)$ in Eq.~(\ref{chi2}) we use the BC03 SSP models computed for the Padova 1994 stellar evolutionary tracks,
the STELIB stellar atlas, and the \citet{chabrier03} IMF (see BC03 for details and references).
To model the galaxy spectrum we can use in Eqs.~(\ref{flamsum}) and (\ref{chi2}) the full age and metallicity resolution available in the BC03
models, i.e. $N_t = 221$ time-steps  and $K_Z= 7$ metallicities.
However, the problem of retrieving physical information from $N_t \times K_Z = 1547$ coefficients presents difficulties.
Some of these difficulties are related to well known degeneracies characteristic of stellar populations \citep[see e.g.][]{conroy13},
with the consequence that more than one SFH will produce the same SED, but the implied galaxy properties
will be different for each of these SFHs (cf. Figure \ref{motiv}).
For this reason, solving the problem of SED fitting goes beyond the solution of a mathematical equation.
The chosen solution must warrant that the physical parameters characterizing the problem galaxy
(e.g., galaxy mass, mean stellar age, dust content, and mean metallicity) are well recovered,
or, at least, the fitting algorithm must provide us with a reasonable estimate of the error on each of these quantities.

In the following we discuss three different non-parametric approaches to solving the SED-fitting problem:
\gaspex, a fixed-basis method;
\tgaspex, an unconstrained basis method; and
\dinbas, a dynamical basis selection method.

\subsection{Three non-parametric methods}\label{threealgs}

\subsubsection{\gaspex: a fixed basis method}

The GAlaxy Spectrum Parameter EXtraction (\gaspex) code is suitable for deriving the SFH of a galaxy using a rather simple SED-fitting procedure\footnote{Preliminary results obtained with this code were presented briefly in \citet{mateu01} and \citet{magris07}, but without a full introduction to the method.}. \gaspex~minimizes $\chi^2$ using {\it a fixed set of n model spectra} (the same for all problem galaxies)
as a basis in Eq.~(\ref{chi2}).
If $m$ is the number of wavelength points in the problem spectrum, we can think of
$F^{obs}_{\lambda_l}/\sigma_{\lambda_l}$ in Eq.~(\ref{chi2}) as the $m$ components of an $1 \times m$ vector $O$.
The subset of $n$ model spectra, $f_{\lambda_l}(t_i,Z_{j})/\sigma_{\lambda_{i}}$, can then be arranged as an $m \times n$ matrix $\mathcal{M}$,
where the number of time steps $t_{i}$ and metallicities $Z_{j}$ used in the fit adds to $n$.
The coefficients $a_{i,j}$ in Eq.~(\ref{chi2}) will then define the $n \times 1$ solution vector $A$.
Minimizing \ch~in Eq.~(\ref{chi2}) is then equivalent to solving for $A$ the following matrix equation

\begin{equation}
O\ =\ \mathcal{M} \times A.
\label{problem}
\end{equation}

\noindent From physical grounds, all components of the solution vector $A$ must be non-negative and
the sum of these components is proportional to the problem galaxy mass.
The Non-Negative Least Squares algorithm (NNLS) developed by \citet{lh74} is especially suited for this purpose.
\gaspex\ uses the NNLS routine to solve Eq.~(\ref{problem}) for $A$ by least squares techniques,
minimizing the squared residuals of the $O =  \mathcal{M}  \times A$ spectral fit,
subject to the conditions $a_{i} \ge 0$ for all $i$, and returns the $n$ components of the vector $A$
that minimize \ch~defined in Eq.~(\ref{chi2}).

An important issue in SED fitting is the degeneracy introduced in the results by using a large number of template SSP spectra as a basis in Eq.~(\ref{chi2}).
Therefore, \textit{one must carefully choose a subset of templates that minimize this degeneracy}.
In Appendix A we explain the procedure used to select an $n=12$ element optimal basis for \gaspex.

\subsubsection{\tgaspex: an unconstrained basis method}

\tgaspex\ is the extension of the \gaspex~algorithm using as a basis the full set  of 221 spectra per metallicity in the BC03 models.
In this case Eq.~(\ref{problem}) is solved using the NNLS routine with $n \le N_t \times K_Z = 221 \times 7$.
The value of $n$ can be reduced if we have some knowledge about the target spectrum.
For instance, for single metallicity fits $n = N_{t} = 221$, in contrast with GASPEX which uses only $n = 12$.

\subsubsection{\dinbas: a dynamical basis selection method}

\citet{mateu09} briefly introduced the basic idea of the non-parametric dynamical basis-selection algorithm (\dinbas)\ for SED fitting and parameter extraction.
In this method, $F^{obs}_\lambda$ in Eq.~(\ref{chi0}) is fitted using $N$ model spectra selected dynamically from the full set available in the stellar population synthesis models (e.g. $N$ out of 221 in the case of the BC03 models for a single metallicity). 
The specific $N$ spectra will vary with the problem galaxy. 
The galaxy physical parameters are then recovered from the weight assigned to each of the $N$ components in the spectral fit.
This method guarantees an optimal solution for a given dimension $N$ of the spectral basis.
This algorithm is quite flexible and easy to implement, but its usefulness remains to be established through internal consistency tests using a large sample of spectra of galaxies of different types.
We denote this code as \dinbasnd\ to indicate the number of $N$ spectra $f_\lambda(t_i,Z_j)$ used in Eq.~(\ref{chi2}).

In the simplest scenario, a galaxy spectrum can be fitted with a single SED $(N = 1)$ by searching for the set of values of
$(a_{i,j},\ t_i,\ Z_j)$ that minimizes \ch\ in Eq.~(\ref{chi2}). 
The values of  $(a_{i,j},\ t_i,\ Z_j)$ are interpreted as the recovered galaxy mass, age, and metallicity, respectively.
 This SSP solution, which we will refer as a 1D solution, is suitable only for early type galaxies and for some star
clusters \citep{cabrera14}.

For $N \ge 2$, we minimize \ch\ in Eq.~(\ref{chi2}) requiring that the $N$ derivatives 
$\frac{\partial}{\partial a_{i}}\chi^{2} = 0$.
This results in a system of $N$ equations with $N$ unknowns that we solve using Cramer's rule for all possible combinations of $N$ model spectra.
The \dinbasnd~solution is then the one with the minimum \ch, subject to the condition $a_{i} \ge 0$ for $i = 2,..., N$.
We will show in the next section that despite its simplicity, this method for $N$ = 2 and 3 is very efficient as well as robust.
The spectral fits are excellent, and the residuals of the recovered physical parameters for the problem galaxies are
less biased than for fixed-age, rigid basis methods.\footnote{In a strict sense, \gaspex\ and \tgaspex\ also select $dynamically$ the $N$
model spectra that minimize \ch\ in Eq.~(\ref{chi2}). In our implementation of these codes, \gaspex\ selects dynamically $N_{G}$ spectra from a fixed set
of $n = 12$ template spectra, the same for all problem galaxies, whereas \tgaspex\ selects dynamically $N_{T}$ spectra from a set of templates as large
as the population synthesis models allow. With the NNLS algorithm we cannot predict or fix in advance the values of $N_{G}$ and $N_{T}$ for a 
problem galaxy. In general, $N_{G}$ will differ from $N_{T}$. On the contrary, the \dinbasnd\ code forces the solution to Eq.~(\ref{chi2}) to include a
number of $ND$ components, requested a priori by the user. \dinbasnd\ provides an exact solution, obtained by inverting the matrix $\mathcal{M}$ in Eq.~(\ref{problem}). The NNLS procedure used by \gaspex\ and \tgaspex\ finds by least squares techniques an approximate solution that minimizes
\ch\ in Eq.~(\ref{chi2}), avoiding the time consuming task of inverting the matrix $\mathcal{M}$.  In the few cases in which $N_{T} = ND$, the selected templates and the set of coefficients $a_{i,j}$ 
may differ for the \tgaspex\ and the \dinbasnd\ solutions.}

\subsection{Treatment of dust extinction and stellar velocity dispersion}

We incorporate the dust extinction and line-of-sight velocity dispersion as model parameters, which we treat in a separate manner given that the model dependency on them is non-linear. 

\subsubsection{Extinction by Dust}\label{tauv}

As indicated in Appendix B, in the \ssag~we use the \citet{cf00} model to
simulate the absorption of starlight by dust in galaxies.
This model treats the effects of dust in star forming galaxies as a two phase process.
In the first phase, young stars are embedded in their native clouds for a short time ($10^7$ yr), and the effects of dust are described by an optical depth
parameter $\tau_{V}^{BC}$.
The second phase corresponds to the absorption in the ambient interstellar medium, and is characterized by an optical depth $\tau_V^{ISM}$.

It is too ambitious to attempt to recover separately the values of $\tau_V^{BC}$ and $\tau_V^{ISM}$ using SED-fitting algorithms, especially for massive data processing. 
Tojeiro et al. (2007) tested VESPA on model spectra which were simulated using the two-phase dust model.
They report a good recovery of $\tau_V^{ISM}$ and a poorer one of $\tau_V^{BC}$.
They also analyzed their sample using a single dust component model, and concluded that the one component model gives acceptable results that are only slightly improved by the two parameter fitting.
\citet{chen12}, using principal component analysis, also estimate dust absorption using just one parameter, although their model spectral library was generated using the two-component \citet{cf00} prescription.

In this paper, we use a one parameter single screen model to describe the effects of dust attenuation in our sample of problem galaxies. 
We adopt the extinction curve from \citet{cf00}: $\tau_\lambda=\tau_V (\lambda/5500$\AA$)^{-0.7}$, 
and look for the minimum value of $\chi^2$ in Eq.~(\ref{chi2}), when the model SED is replaced by the respective reddened SED:  
$f^{redd}_\lambda =  f_\lambda \exp(-\tau_\lambda)$.
In practice, for each galaxy in the sample and for each algorithm described above,  we repeat the fits using a fixed set of values of $\tau_V$, including $\tau_V = 0$.
The value of $\tau_V$ for which \ch~attains an absolute minimum for the specific algorithm is then adopted as its best estimate of $\tau_V$, and is reported in terms of the $V$-band extinction, $A_V = 1.086 \tau_V$.

\subsubsection{Line of sight velocity dispersion}

The same brute-force procedure used to find $A_V$ is followed to determine the value of $\sigma_v$, the velocity 
dispersion characterizing the line-of-sight stellar motions in the problem galaxy.
For each galaxy and for each algorithm, we repeat the fits convolving the model SEDs with a Gaussian kernel,
increasing $\sigma_v$ from 0 to 300 km s$^{-1}$ in steps of 10 km s$^{-1}$.
Our estimate of $\sigma_v$ is taken as the value that produces the minimum \ch.
Given the quality of the spectral fits achieved by the different codes, the adopted value of $\sigma_v$ is practically independent
of the algorithm within a precision of 10 km s$^{-1}$, which suffices for our purposes.
For practical reasons the fits reported in this paper were performed with the value of $\sigma_v$ determined with \gaspex.

\section{Testing different algorithms}\label{tests}

\begin{figure}
\plotone{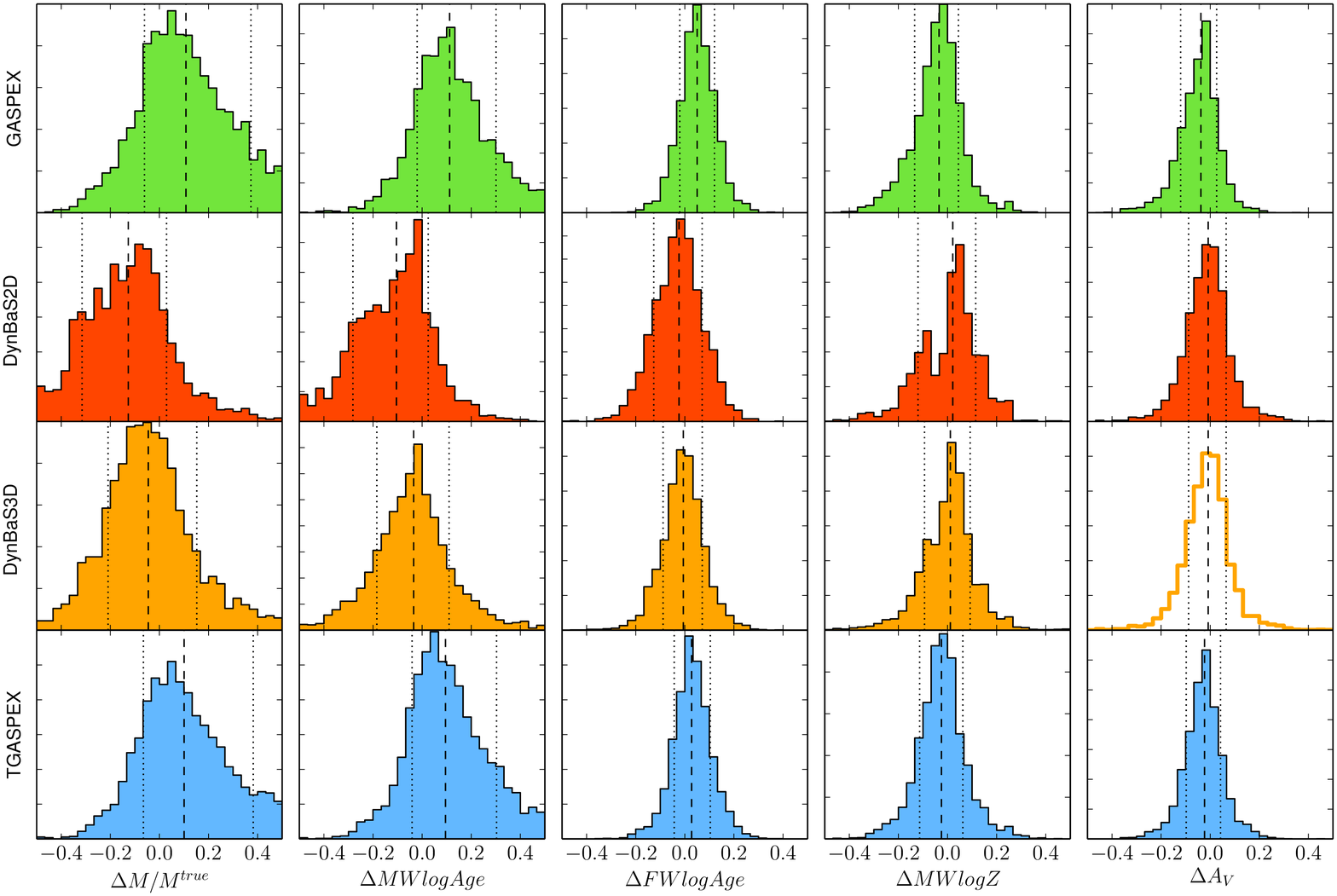}
\plotone{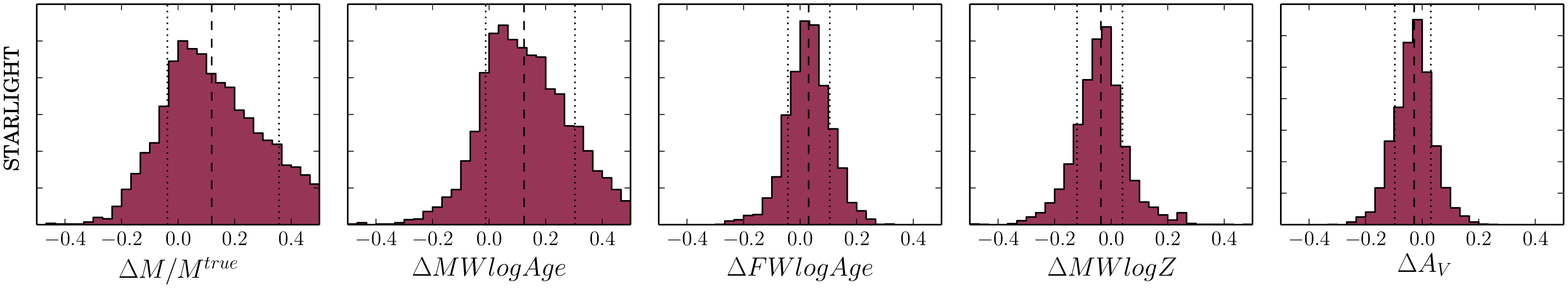}
\caption{Distributions of residuals obtained with \gaspex, \dinbasdos, \dinbastres, \tgaspex, and \starlight~for the total mass $M$, mass-weighted log-age $MWlogAge$, flux-weighted log-age  $FWlogAge$, mass-weighted log-metallicity $MWlogZ$, and V-band extinction $A_V$ for mock galaxy spectra with SNR$r$=20.  
The median and the 16th and 84th percentiles are indicated with dashed and dotted lines, respectively. The distribution of $\Delta A_V$ in the \dinbastres~panel corresponds to the one obtained with \dinbasdos~(see text for details). All distributions are normalized to unit area (the vertical scale in each panel is different).
}
\label{hist_test1}
\end{figure}

\begin{figure}
\plotone{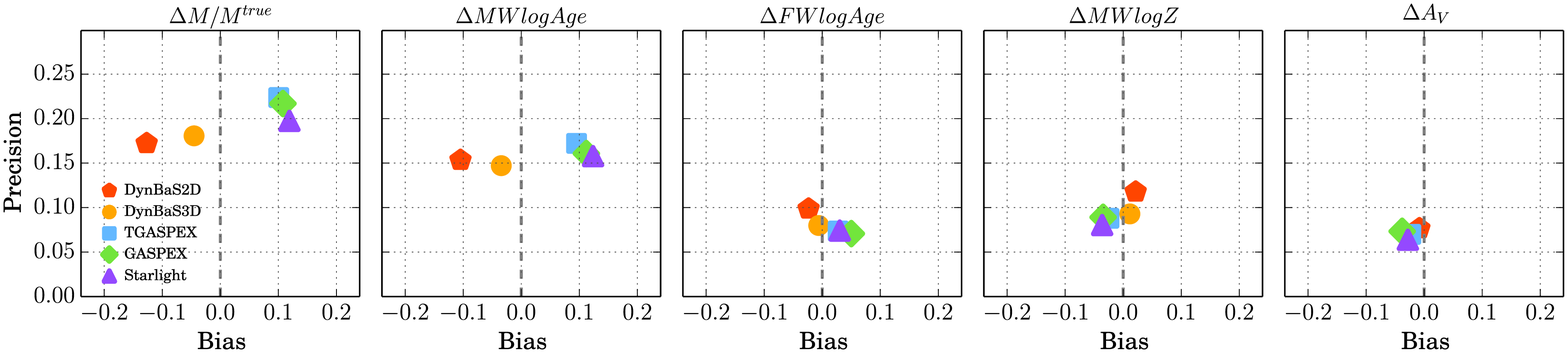}
\plotone{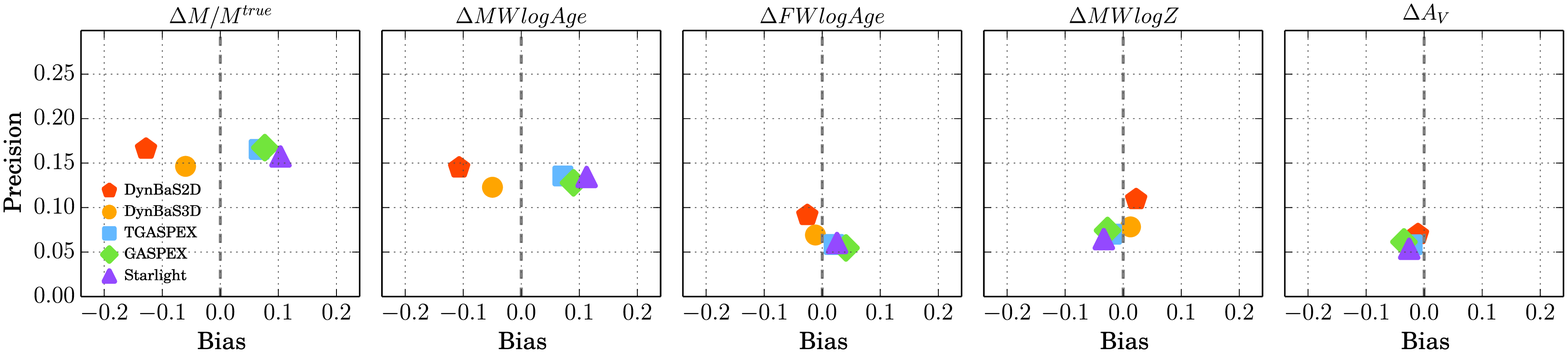}
\caption{Precision vs.~bias obtained with \gaspex, \dinbasdos, \dinbastres, \tgaspex, and \starlight~for the recovery of the total mass $M$, mass-weighted
log-age $MWlogAge$, flux-weighted log-age  $FWlogAge$, mass-weighted log-metallicity $MWlogZ$, and V-band extinction $A_V$
(panels from left to right). The bias and precision have been defined respectively as the median of the residuals and the half distance between the 16th and 84th percentile of residuals (see text for discussion).
The top panels correspond to SNR$r$=20 mock galaxy spectra and the bottom ones to SNR$r$=30.}
\label{biaserr}
\end{figure}

\begin{figure}
\plotone{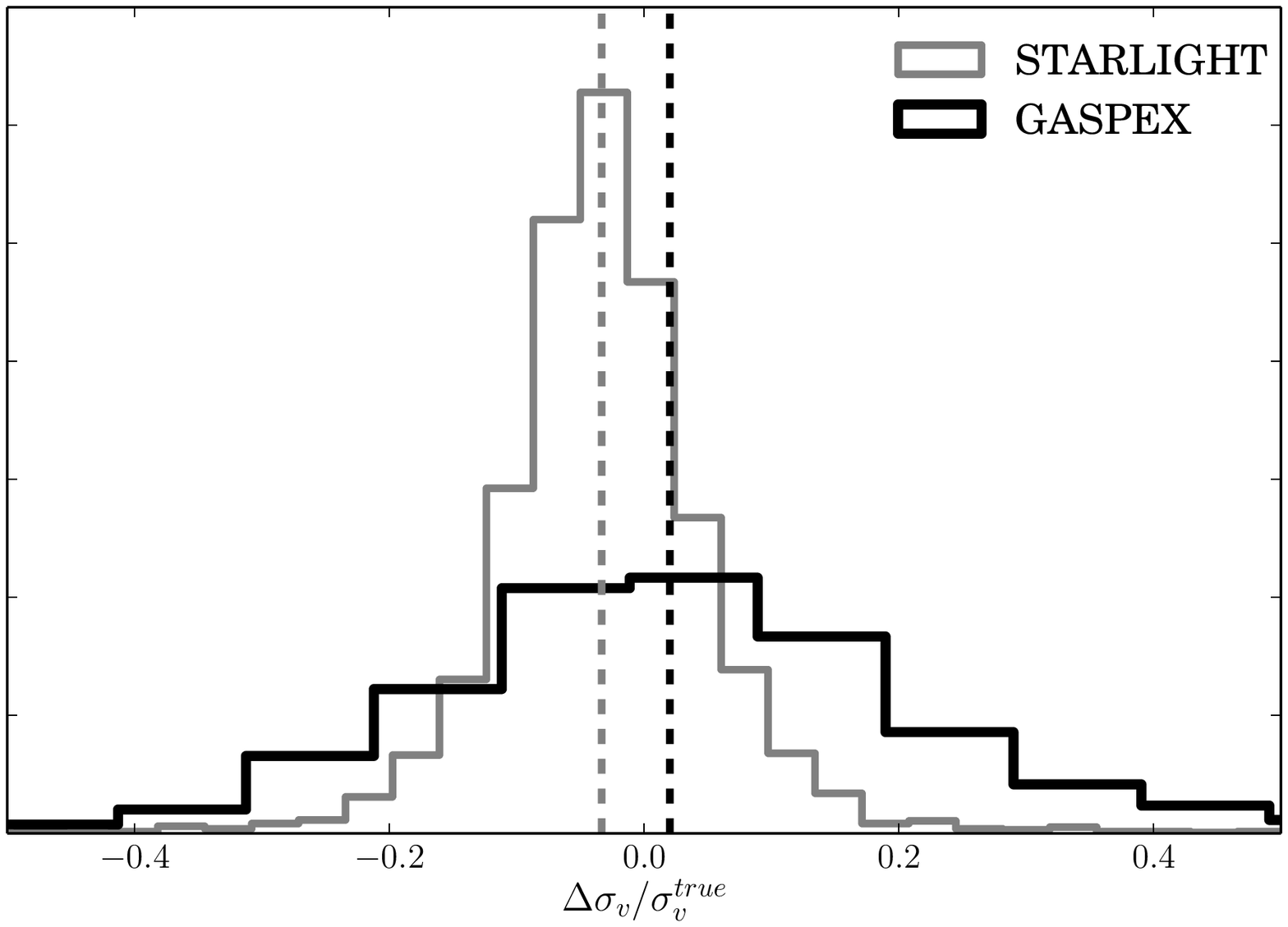}
   \caption{Distributions of residuals in the determination of the stellar velocity dispersion $\sigma_v$ using \gaspex\ and \starlight\ on mock spectra with SNR$r$=20. 
   The vertical dashed lines indicate the median of residuals for each model.
   The distributions are normalized to unit area.
   }
  \label{sigma}
\end{figure}

A subset of 120 SEDs from the 
\ssag\ is used to evaluate the general performance of the different algorithms described in \S 2 in recovering the physical properties of galaxies.
The selected galaxies follow the bimodal distribution in the $(u-g)$ vs. $(g-r)$ color-color plane
defined by SDSS galaxies up to redshift $z=0.03$ shown in Figure \ref{sampled} \citep{strateva01}.
We limit the synthetic SEDs to the spectral range from 3300 to 8000 \AA, similar to the SDSS rest-frame spectral coverage for nearby galaxies, 
perfectly suited for our purposes.
We add random gaussian noise to the flux in each of the selected SEDs, with a wavelength dependent standard deviation
derived from the average of the error-array of spectroscopic data of SDSS DR7, scaled to a fixed value of the SNR in the $r$ band (SNR$r$). 
This process is repeated 30 times for each galaxy in the sample, for a total of 3,600 problem spectra.
To test the dependence of the results on the quality of the spectra, we use SNR$r$ = 20 and 30.

\begin{deluxetable}{llrrrcrrrrc}
\tabletypesize{\scriptsize}
\tablecolumns{1}
\tablewidth{0pt}
\tablecaption{{Statistics of the residual distributions}}
\tablehead{
\colhead{}  & \colhead{} & \multicolumn{4}{c}{SNR$r$=20} & \colhead{} & \multicolumn{4}{c}{SNR$r$=30} \\
\cline{3-6} \cline{8-11} \\
\colhead{Parameter} & \colhead{Method} & \colhead{Median\tablenotemark{a}} & \colhead{P16} & \colhead{P84} & \colhead{Precision\tablenotemark{b}}&
\colhead{} &
\colhead{Median\tablenotemark{a}} & \colhead{P16} & \colhead{P84} & \colhead{Precision\tablenotemark{b}}
}
\startdata
\multirow{5}{*}{$\Delta M/M^{true}$}          & \gaspex     &   0.108 &   -0.061 &    0.373 &    0.217 & &  0.077 &   -0.051 &    0.283 &    0.167 \\ 
                                              & \dinbasdos  &  -0.127 &   -0.315 &    0.029 &    0.172 & & -0.128 &   -0.315 &    0.017 &    0.166 \\ 
                                              & \dinbastres &  -0.045 &   -0.209 &    0.152 &    0.181 & & -0.060 &   -0.202 &    0.090 &    0.146 \\ 
                                              & \tgaspex    &   0.101 &   -0.065 &    0.382 &    0.224 & &  0.067 &   -0.056 &    0.275 &    0.165 \\ 
                                              & \starlight  &   0.119 &   -0.038 &    0.357 &    0.197 & &  0.104 &   -0.021 &    0.293 &    0.157 \\
\hline
\multirow{5}{*}{$\Delta MWlogAge$}            & \gaspex     &   0.112 &   -0.020 &    0.301 &    0.161 & &  0.090 &   -0.010 &    0.245 &    0.128 \\ 
                                              & \dinbasdos  &  -0.105 &   -0.282 &    0.025 &    0.153 & & -0.107 &   -0.282 &    0.007 &    0.145 \\ 
                                              & \dinbastres &  -0.035 &   -0.184 &    0.110 &    0.147 & & -0.050 &   -0.182 &    0.063 &    0.123 \\ 
                                              & \tgaspex    &   0.095 &   -0.041 &    0.303 &    0.172 & &  0.072 &   -0.033 &    0.238 &    0.136 \\ 
                                              & \starlight  &   0.124 &   -0.012 &    0.304 &    0.158 & &  0.113 &    0.001 &    0.269 &    0.134 \\ 
\hline
\multirow{5}{*}{$\Delta FWlogAge$}            & \gaspex     &   0.050 &   -0.021 &    0.121 &    0.071 & &  0.041 &   -0.013 &    0.096 &    0.055 \\ 
                                              & \dinbasdos  &  -0.024 &   -0.127 &    0.071 &    0.099 & & -0.026 &   -0.121 &    0.061 &    0.091 \\ 
                                              & \dinbastres &  -0.006 &   -0.089 &    0.071 &    0.080 & & -0.012 &   -0.083 &    0.055 &    0.069 \\ 
                                              & \tgaspex    &   0.028 &   -0.043 &    0.104 &    0.074 & &  0.020 &   -0.036 &    0.081 &    0.058 \\ 
                                              & \starlight  &   0.030 &   -0.043 &    0.105 &    0.074 & &  0.025 &   -0.036 &    0.084 &    0.060 \\ 
\hline
\multirow{5}{*}{$\Delta MWlog(Z)$}            & \gaspex     &  -0.035 &   -0.134 &    0.044 &    0.089 & & -0.027 &   -0.115 &    0.033 &    0.074 \\ 
                                              & \dinbasdos  &   0.021 &   -0.120 &    0.115 &    0.118 & &  0.022 &   -0.105 &    0.113 &    0.109 \\ 
                                              & \dinbastres &   0.012 &   -0.094 &    0.092 &    0.093 & &  0.013 &   -0.077 &    0.080 &    0.078 \\ 
                                              & \tgaspex    &  -0.025 &   -0.113 &    0.062 &    0.088 & & -0.020 &   -0.093 &    0.047 &    0.070 \\ 
                                              & \starlight  &  -0.036 &   -0.121 &    0.040 &    0.080 & & -0.033 &   -0.102 &    0.027 &    0.065 \\ 
\hline   
\multirow{5}{*}{$\Delta A_V$}                 & \gaspex     &  -0.038 &   -0.121 &    0.026 &    0.073 & & -0.035 &   -0.111 &    0.012 &    0.061 \\ 
                                              & \dinbasdos  &  -0.008 &   -0.088 &    0.065 &    0.077 & & -0.011 &   -0.085 &    0.055 &    0.070 \\ 
                                              & \tgaspex    &  -0.023 &   -0.098 &    0.042 &    0.070 & & -0.021 &   -0.086 &    0.030 &    0.058 \\ 
                                              & \starlight  &  -0.028 &   -0.097 &    0.031 &    0.064 & & -0.026 &   -0.082 &    0.025 &    0.054 \\ 
\hline
\multirow{2}{*}{$\Delta \sigma_v/\sigma_v^{true}$}& \gaspex     &   0.020 &   -0.148 &    0.211 &    0.180 & &  0.018 &   -0.099 &    0.161 &    0.130 \\ 
                                              & \starlight  &  -0.033 &   -0.099 &    0.036 &    0.067 & & -0.027 &   -0.074 &    0.020 &    0.047 \\ 
\enddata
\tablenotetext{a}{We use the median as a measure of the bias of the distribution.}
\tablenotetext{b}{We define precision as (P84$-$P16)/2.}

\label{t:residuals}
\end{deluxetable}

\subsection{Recovering Physical Parameters: Residual Distributions}

We fit each problem SED using the \dinbasdos, \dinbastres, \gaspex, \tgaspex, and \starlight\ algorithms,
using the BC03 SSP models for $Z = (0.2, 0.4, 1, 2.5) \times Z_\odot$ with a maximum model age of 13.75 Gyr as basic ingredients.
For \starlight\ we use the updated 39 age basis proposed by \citet{cidfernandes14}.
From each fit we recover the following galaxy properties:
total mass, $M$; 
mass-weighted log-age, $MWlogAge$; 
flux-weighted log-age\footnote{The $r$-band flux is used as weight to compute the flux-weighted log-age}, $FWlogAge$;
mass-weighted log-metallicity, $MWlogZ$; 
$V$-band extinction, $A_V$; and
stellar velocity dispersion, $\sigma_v$,
and build the distributions of the residuals\footnote{
Residuals are defined as $\Delta x = (x - x^{true})$, where $x$ is the recovered value of a galaxy property, and $x^{true}$ its true value.
The true values are determined from the galaxy SFH and the population synthesis models for all the properties except for $A_{V}$.
For the later we use $A_V^{true} = V^{redd}-V$, where $V^{redd}$ and $V$ represent the $V$-band magnitude of the reddened and
unreddened model galaxy computed for the same SFH, respectively.
For the total mass and velocity dispersion, we use the fractional residuals $\Delta M/M^{true}$ and $\Delta\sigma_v/\sigma_v^{true}$ respectively.} shown in Figures \ref{hist_test1}.
The median, 16th and 84th percentile\footnote{The 16th and 84th percentile are the values below which 16 and 84 percent of the residuals may be found. For a normal distribution they correspond approximately to $-1\sigma$ and $+1\sigma$, respectively.}
(referred hereafter as P16 and P84, respectively) of these distributions are listed in Table \ref{t:residuals}, and are indicated in the
figures by the dashed and dotted vertical lines, respectively.
In the determination of $M,\ MWlogAge,\ FWlogAge$, and $MWlogZ$ we use the best estimates of $A_{V}$ and $\sigma_v$ 
obtained as described below. 
 
A quantitative assessment of the overall performance of the different fitting methods on our sample is summarized in Figure \ref{biaserr}. We use the median of the residual distribution as a measure of the $bias$,
and its width, defined here as $(P84 - P16)/2$, as an indicator of the \emph{precision}
in the determination of the corresponding parameter, both measured globally over the whole galaxy sample.
Below we discuss in detail the implications of Figures \ref{hist_test1} and \ref{biaserr}, and Table \ref{t:residuals} for the case SNR$r$$=$20 and 30. 

\subsubsection{Mass}\label{s:mass}

For spectra with SNR$r$=30, all methods recover the stellar mass with essentially the same precision, $\lesssim17$\%, but differ in the bias (see Table \ref{t:residuals}).
\dinbastres\ tends to underestimate $M$ by $\sim6$\%, and \gaspex~and \tgaspex, on average, tend to overestimate $M$ by a similar amount
(bias $\sim 7$\%).
\dinbasdos\ shows the most biased estimation of $M$ (bias $\sim 13$\%) and an extended asymmetrical distribution of residuals.
For lower quality spectra with SNR$r$$=20$, we find larger differences in the bias reported for each method. 
In this case, \dinbastres\ provides the less biased estimate of the stellar mass $M$ (negative bias of $5$\%), while \gaspex\ and \tgaspex\ recover $M$ with a bias of $\sim10$\%. 
Results in Table \ref{t:residuals} show that the residual in the determination of $M$ for \gaspex\ and \tgaspex\ are less biased for higher SNR spectra, while \dinbastres\ shows the inverse trend.
The \starlight~residuals are very similar to the \gaspex\ and \tgaspex\ residuals, with a distribution slightly more biased but also slightly more precise.

\subsubsection{Mass-weighted log-Age}\label{s:mlogage}

Even though the precision for all methods is comparable ($0.15-0.17$ dex and $0.12-0.15$ dex, for SNR$r$=20 and 30, respectively), the bias in the residuals of $MWlogAge$\ shows the same trend as the stellar mass $M$ residuals: \dinbasdos\ and \dinbastres\ tend to underestimate $MWlogAge$ (bias $\sim-0.11$ and $\lesssim-0.05$ dex, respectively) with a weak dependence on the quality of the spectra,  while \gaspex\ and \tgaspex\ overestimate $MWlogAge$ with bias $\lesssim0.11$ and $\lesssim0.09$, for SNR$r$=20 and 30, respectively.
As for the mass $M$, the $MWlogAge$ determination improves with SNR for \gaspex\ and \tgaspex, while it worsens for \dinbasdos\ and \dinbastres.
The results obtained with \starlight\ are practically identical to those of \gaspex\ and \tgaspex.

\subsubsection{Flux-weighted log-Age}\label{s:flogage}

\dinbastres\ and \tgaspex\ produce almost unbiased estimates of $FWlogAge$, with a bias $\lesssim-0.01$ and $\lesssim0.03$ dex, respectively.
All methods, \dinbasdos, \dinbastres, \gaspex\ and \tgaspex\ provide equally precise estimates of $FWlogAge$ (precision $\leq 0.1$ dex), 
practically independent of the SNR. 
\starlight\ gives the same statistical estimation of $FWlogAge$ as \tgaspex.

\subsubsection{Mass-weighted log-Metallicity}\label{s:z}

The typical precision achieved in the recovery of $MWlogZ$ is quite good, and the distributions of residuals are almost independent of 
the SNR of the target spectrum.
The \dinbastres, \gaspex, \tgaspex\, and \starlight\ residuals show very little bias, $0.01$, $-0.04$, $-0.03$, and $-0.04$ dex, respectively, 
with a precise estimate of $MWlogZ$ $\leq 0.09$ dex. 
For \dinbasdos\ the $MWlogZ$ residual distribution is bimodal, indicating a limitation of this method to determine a galaxy metallicity from its spectrum.

\subsubsection{Extinction}\label{s:av}

We have used \dinbasdos, \gaspex, and \tgaspex\ to estimate $A_V$ by brute-force as indicated in \S \ref{tauv}.
\dinbasdos, \gaspex, \tgaspex\, and \starlight\ provide practically unbiased estimates of $A_{V}$, bias between $-0.01$ and $-0.04$, with precision $\leq 0.08$.
We did not perform the brute-force determination of $A_V$ for \dinbastres, since
for a reduced subsample of galaxies we recovered the same value of $A_V$ as for \dinbasdos.
The value of $A_V$ does not change significantly with SNR.

\subsubsection{Velocity Dispersion}\label{s:lossm}

Figure \ref{sigma} shows as histograms the distributions of fractional residuals in the derivation of $\sigma_v$ for \gaspex\ and \starlight.
The velocity dispersion recovered by \gaspex\ is almost unbiased (bias 0.02), with a precision of 18 and 13\% for SNR$r$=20 and 30, respectively.
 \starlight\ produces a narrower distribution of fractional residuals with precision $= 0.07$ and 0.05 for SNR$r$$=$ 20 and 30, respectively, but underestimates $\sigma_v$ slightly (bias $= -0.03$).

\subsubsection{Residuals as a function of galaxy color}

\begin{figure}
\plotone{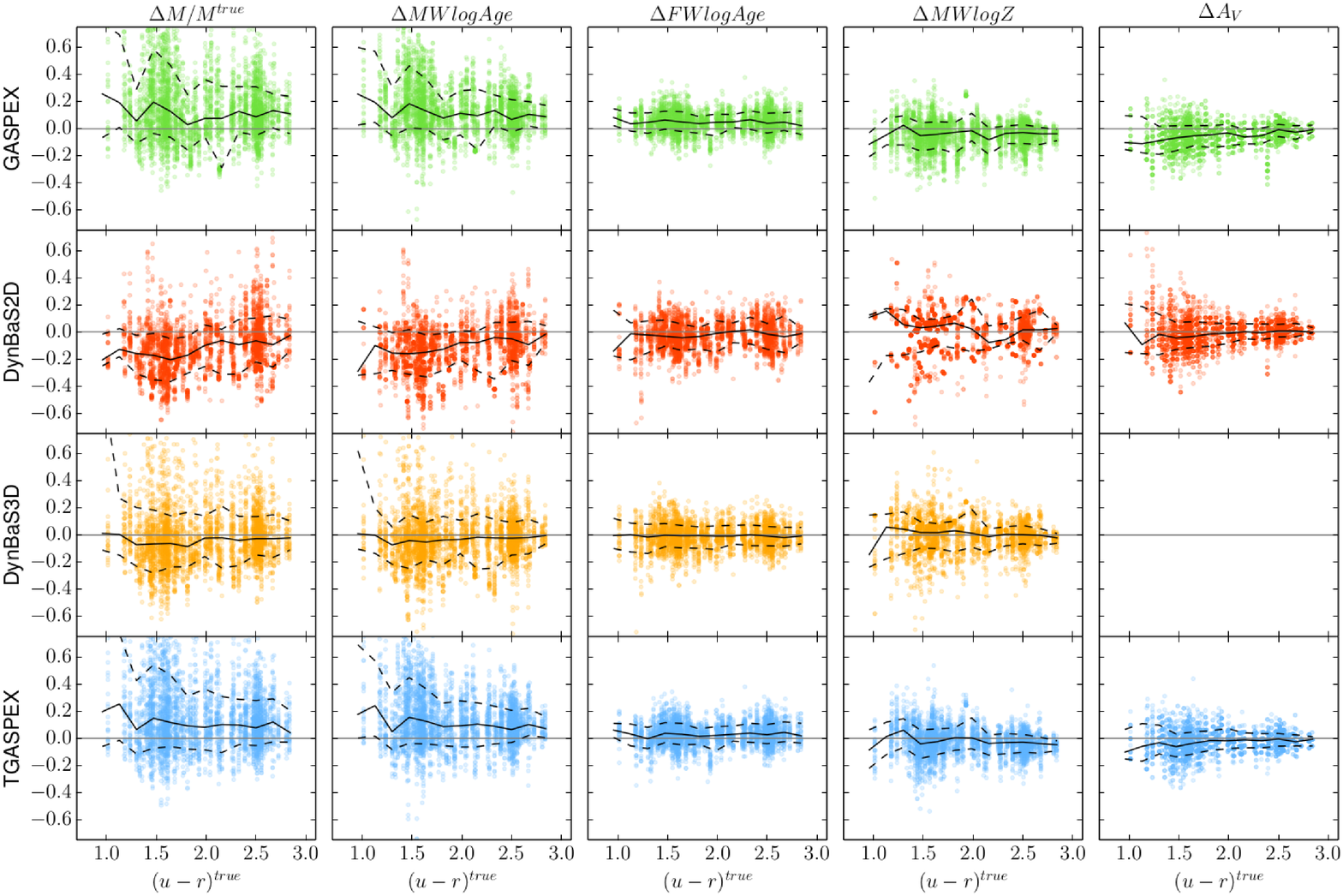}
\plotone{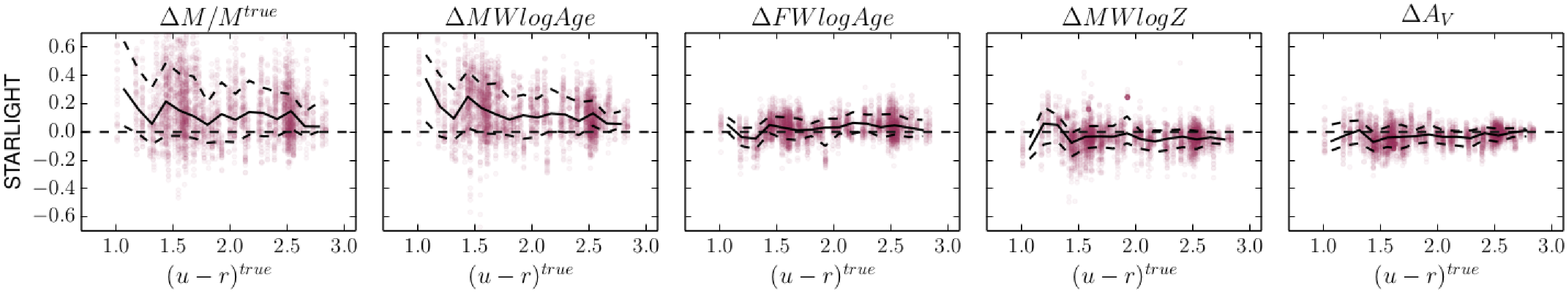}
   \caption{Residuals as a function of $\sur$ color for SSAG spectra with SNR$r$=20. The median and the 16th-84th percentile of the residual distributions are indicated with solid and dashed lines, respectively.
   }
\label{residuos_ur}
\end{figure}

\begin{figure}
\plotone{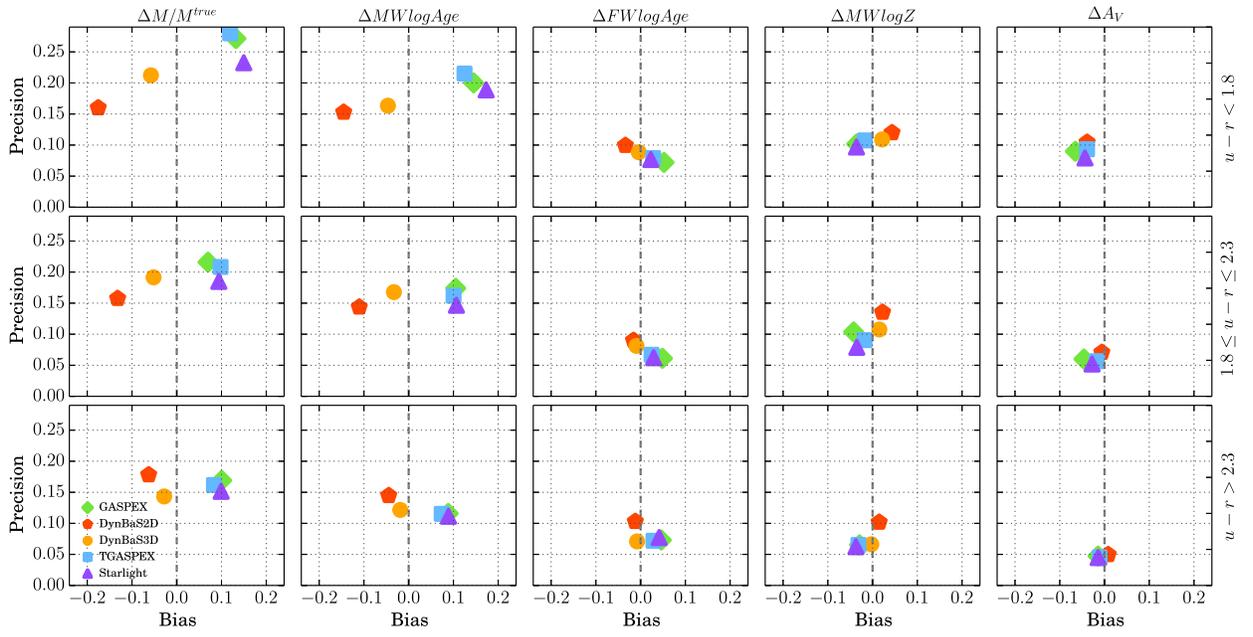}
   \caption{Same as top panels of Figure \ref{biaserr} (SNR$r$$=$20), but for galaxies in three different color bins: $\sur<1.8$ (\emph{top}), 
   $1.8\leqslant \sur\leqslant 2.3$ (\emph{middle}), and $\sur>2.3$ (\emph{bottom}).
   }
\label{biaserr123}
\end{figure}

The $M$, $MWlogAge$, $FWlogAge$, $MWlogZ$, and $A_V$ residuals as a function of galaxy $\sur$ color are shown in Figure \ref{residuos_ur} for the 3,600 spectra in our sample with SNR$r$=20. 
Figure \ref{biaserr123} provides a graphical summary of the residual distributions of Figure \ref{residuos_ur},
showing the bias and precision obtained with all methods for galaxies in three $\sur$ color bins (bluer galaxies on top panels).
This figure clearly shows that for all methods the bias decreases (the distribution of points shrinks towards zero in the horizontal direction), 
and the precision improves (points move towards zero in the vertical direction), when going from the bluer to the redder bins.
Overall, \dinbastres\ provides the least biased estimate of all these properties for galaxies of any $\sur$ color.
The values of $M$ and $MWlogAge$ recovered by \gaspex, \tgaspex, and \starlight\ show a systematic bias,
being larger than the respective true value for galaxies of any color.
Precision does not vary significantly between the different methods as a function of galaxy color. The same trend is observed for mock spectra with SNR$r$=30.

\subsubsection{Residuals and the chemical evolution in galaxies}

\begin{figure}
\plotone{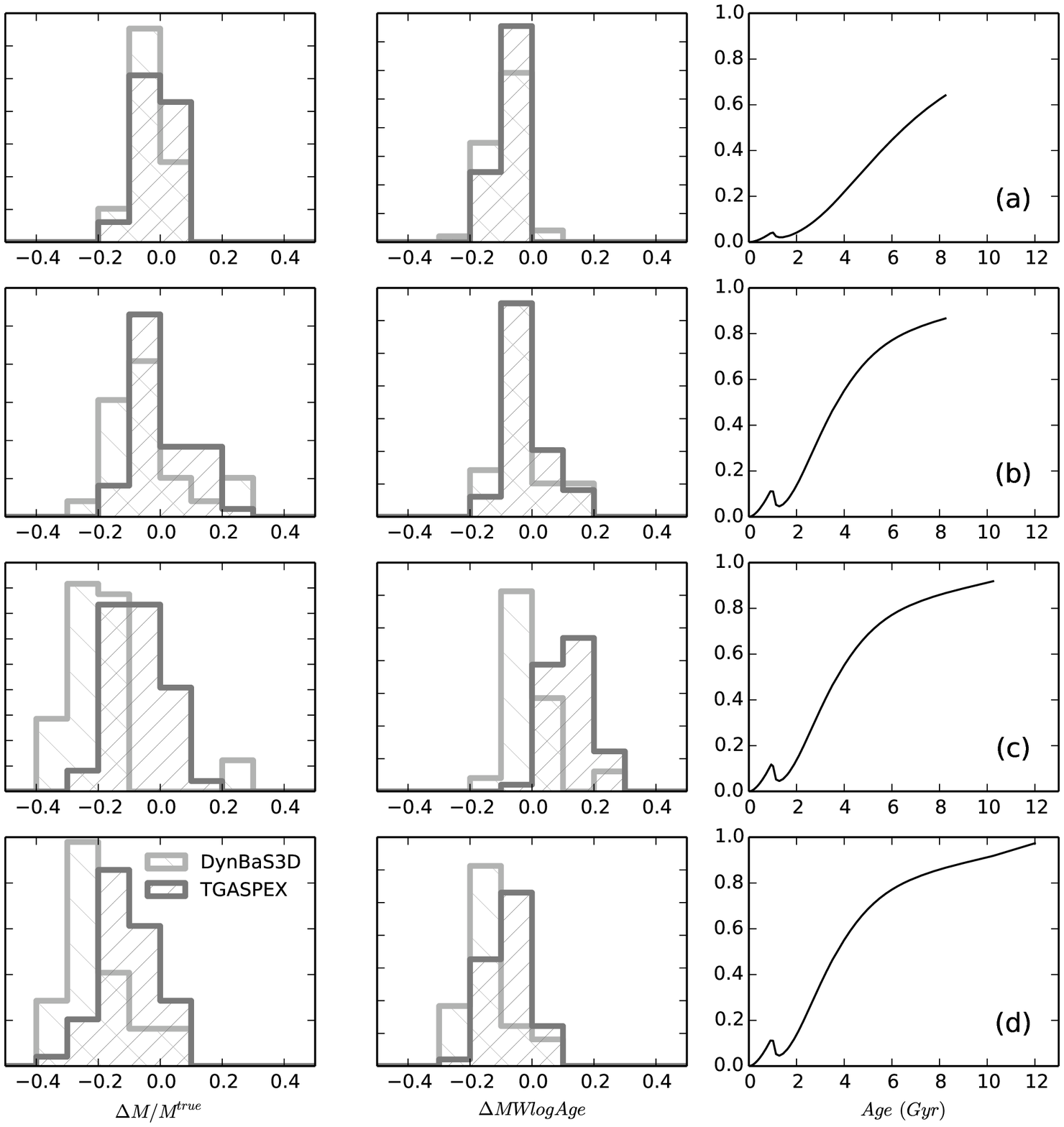}
   \caption{Distribution of residuals obtained with \dinbastres\ and \tgaspex\ for the total mass $M$ and the mass-weighted log-age $MWlogAge$ (left and middle panel) for four synthetic galaxies with metallicity $Z$ evolving in time as shown in the right hand side panels assuming SNR$r$=20.
     }
\label{zt_4gals}
\end{figure}

To assess the effect of the metallicity evolution in the recovered physical parameters, we show in Figure \ref{zt_4gals} 
the residuals in the $M$ and $MWlogAge$ determination using \dinbastres\ and \tgaspex\ as representative of a dynamical basis-selection and an unconstrained basis method, for four different synthetic galaxies with SNR$r$=20. 
The metallicity in these galaxies evolves in time as indicated in the respective right hand side panel of Figure \ref{zt_4gals}.
For clarity, we show only the residuals of the mass related physical quantities, the most sensitive ones to the fitting method,
as shown in the previous sections. 
For \dinbastres\ and \tgaspex\ the residual distributions are very similar for the four galaxies, especially for galaxies in row (a) and (b). 
For galaxy (c), $M$ is clearly underestimated by \dinbastres, and the \tgaspex\ residual distribution is nearly unbiased, 
while $MWlogAge$ is underestimated by \dinbastres\ and overestimated by \tgaspex.  
Finally, for galaxy (d), $M$ and $MWlogAge$ are underestimated by both \dinbastres\ and \tgaspex. 
The mean metallicity $MWlogZ$ is not well recovered by any of the methods and is not shown in Figure \ref{zt_4gals}.  The conclusions do not depend on the explored values of SNR.
These examples illustrate that the residual distributions are not dramatically different for the two methods. 
An analysis of their behaviour as a function of the chemical evolution of each galaxy requires a much more detailed analysis, 
which is out of the scope of this work.

\subsubsection{Summary}

\dinbastres\ shows a slightly better overall performance in the recovery of physical parameters of galaxies with SNR$r=20$. The differences between methods are more notorious in the determination of $M$ and $MWlogAge$, and decrease when SNR$r$ increases from 20 to 30. 
Nevertheless, it should be pointed out that the precision obtained with
\gaspex, \dinbasdos, \dinbastres, \tgaspex~and \starlight\ are comparable for the six parameters explored above.
The results obtained for $M$, $MWlogAge$, $FWlogAge$, $MWlogZ$, $A_V$ and $\sigma_v$ show very minor differences for the different algorithms, and can be considered equivalent.
For a reduced set of mock galaxies with time evolving metallicity, \dinbastres~and \tgaspex\ provide an analogous determination of $M$ and $MWlogAge$.

\section{Discussion and Conclusions}

The goal of SED fitting is to determine from galaxy spectra the physical parameters characterizing
stellar populations in galaxies.
Some of these parameters dictate the past and present evolution of galaxies.
We have discussed the results of blind tests on SED fitting using the 
\gaspex, \dinbasdos, \dinbastres, and \tgaspex~algorithms developed by the authors, and the public code
\starlight~\citep{cidfernandes05, cidfernandes14}, applied to a sample of 120 \ssag~spectra selected to
cover the range of galaxy spectra seen in the Universe.
\emph{All these methods provide roughly equally good spectral fits, but show differences when 
recovering physical parameters.}
Our blind tests allow us to report a quantitative assessment of the precision (error) and bias in the determination of $M$, $MWlogAge$, $FWlogAge$, $MWlogZ$, $A_V$, and $\sigma_v$, using the \gaspex, \dinbasdos, \dinbastres, \tgaspex, and \starlight\ codes.
The values reported in Table \ref{t:residuals} should be used to assign lower limits to the errors 
when spectra of observed galaxies are fitted with any of these algorithms. 

\dinbastres~(dynamically selected spectral basis with three components) shows the best performance,
producing, in general, the least biased determination of stellar mass, age, and metallicity, for mock galaxies with SNR$r$ = 20,
with the same precision as \dinbasdos~(dynamically selected spectral basis with two components), 
\gaspex~(fixed 12-dimensional basis per metallicity), and \tgaspex~(unconstrained basis). 
The visual extinction $A_V$ was recovered only with \dinbasdos, \gaspex, and \tgaspex, with no significant differences in the results.
The determination of the physical parameters provided by all methods is more precise, in general, for SNR$r$=30 than for SNR$r$=20.
\gaspex, \tgaspex\ and \starlight\ produce less biased determinations of $M$, $MWlogAge$, and $FWlogAge$ when SNR$r$ is increased from 20 to 30,
while \dinbasdos\ and \dinbastres\ tend to give more biased results. 

\starlight, on average, overestimates mass and mass-weighted log-age, and shows roughly the same dispersion in the residuals as the other codes. The performance of \gaspex, \tgaspex, and \starlight~is statistically comparable.

\dinbastres~is a powerful and easy to implement algorithm to fit spectra and recover galaxy parameters.
The three stellar populations obtained as the \dinbastres~solution represent the relevant events in the SFH of a galaxy, and
can be used for a statistical estimation of this SFH over the galaxy lifetime.
This will be addressed in an upcoming paper (Mateu, Magris \& Bruzual, in preparation).
In particular, \dinbas~provides a good description of the SFH in stellar systems with isolated episodes of star formation,
which are usually blurred when using other SED-fitting methods \citep{cabrera14}.
If a more detailed estimate of the SFH is desired, \tgaspex~is a more suitable choice.
 
A popular choice in SED fitting is to assume that the SFR declines or increases exponentially with time
\citep{maraston10, hansson12, lee09, pforr12, mitchell13}. 
This may introduce strong biases in the determination of the galaxy stellar mass and age when the true SFR is not well described by
the chosen parametrization, as indicated by \citet{lee09}, \citet{pforr12}, and \citet{mitchell13}.
Such biases are naturally diminished by using non-parametric algorithms like \dinbasdos, \dinbastres, \gaspex, and
\tgaspex, improving the accuracy of parameter determination. 

As a last caveat, we must mention that \emph{the uncertainties in the physical parameters derived in this paper should be considered as lower limits to the errors in
parameter estimation}. Our tests were conducted on synthetic galaxy spectra assembled from the same evolutionary models used to fit the target galaxies.
\emph{Larger errors are expected when real galaxy spectra are fitted, or when an independent set of models is used to fit our \ssag~sample of mock galaxies}.
 
\section*{Acknowledgments}

We thank Drs. L. Aguilar and N. Bastian for their careful reading and comments on a preliminary version of this paper.
GBA acknowledges support for this work from the National Autonomous University of M\'exico (UNAM), through grants IA102311 and IB102212-RR182212.
CM acknowledges the support from the postdoctoral Fellowship of DGAPA-UNAM, M\'exico.
GMC, CM, AMN, and ICZ acknowledge the hospitality of the UNAM Centro de Radioastronom\'ia y Astrof\'isica during part of this investigation.

\appendix

\section{Selecting the optimal basis}

\begin{figure}
\plotone{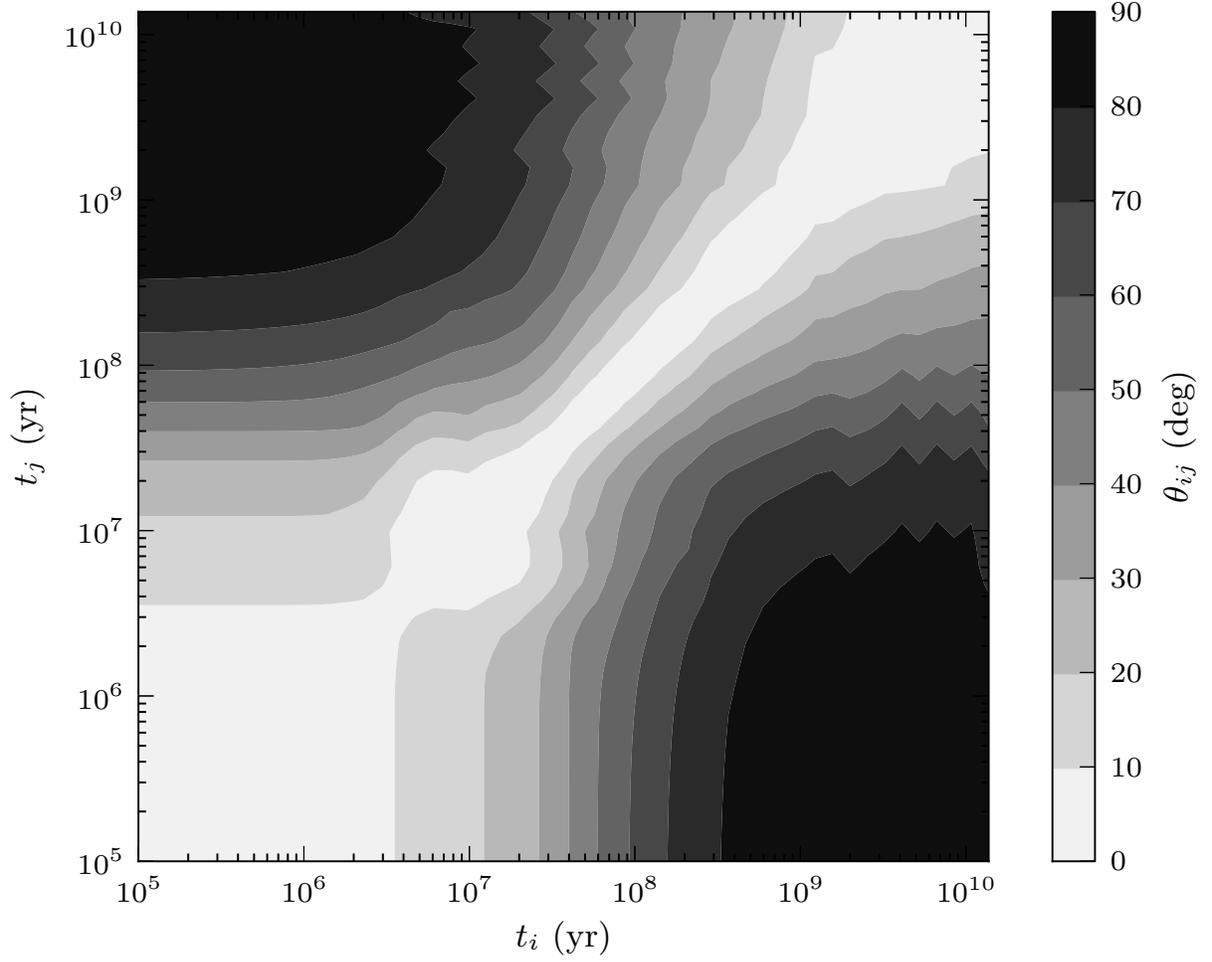}
  \caption{The figure shows the angle $\theta_{ij}$ (Eq. \ref{dp1}) between spectra of different ages in the $Z_\odot$ BC03 model.
  Note that the angle is larger between young and old spectra (upper right and lower left corners), and is small between spectra of similar ages (diagonal pattern).}
 \label{angle}
\end{figure}

We can think of the matrix $\mathcal{M}$ in Eq. \ref{problem} as a basis of an $n$-dimensional vector space.
We are interested in finding in this space an optimal basis of dimension $< n$, which is as orthogonal as possible.
Although conceptually simple, in practice most of the spectra in SSP models are far from orthogonal.
To illustrate this point, we assume that the
$N_{t}$ spectra $f_{i} \equiv f_\lambda(\lambda,t_i) \equiv f_{\lambda}(\lambda,t_{i},Z)$ 
available in a BC03 model (cf. Eq. \ref{flamsum}) define a $N_{t}$-dimensional vector space.
The `angle' between vectors $f_{i}$ and $f_{j}$ is obtained from the dot product $\theta_{ij}$ given by
\begin{equation}
\theta_{ij} = \cos^{-1} \Big(    \frac {\mbox{f}_i \cdot \mbox{f}_j} {\|\mbox{f}_i\| \|\mbox{f}_j\|}   \Big), 
\label{dp1}
\end{equation}
where
\begin{equation}
f_i \cdot f_j = \ \sum_{\lambda = \lambda_1}^{\lambda_m} \  f_\lambda(t_i)  f_\lambda(t_j).
\label{dp2}
\end{equation}
The sum in Eq. \ref{dp2} extends over all wavelength points in the model spectra.
We show in Figure~\ref{angle} the angle between all possible pairs $(i,j)$ of model spectra
in the $Z_{\odot}$ BC03 model.
From this figure it is clear that only vectors with large differences in age are strictly orthogonal. 

To select an optimal basis, that excludes nearly parallel spectra, 
we start by defining a 2-dimensional basis choosing a young ($1.78\times10^{5}$ yr) and an old (13.75 Gyr) spectrum. 
We then fit the remaining $N_{t}-2$ spectra in the model with a linear combination of this 2 element basis.
Using $S$, the sum of quadratic residuals, as a goodness-of-fit indicator, we select the spectrum with the highest $S$ (worse fit) and include it in the basis. We repeat this procedure, adding at each step a new spectrum to the previous basis. 

\begin{figure}
\plotone{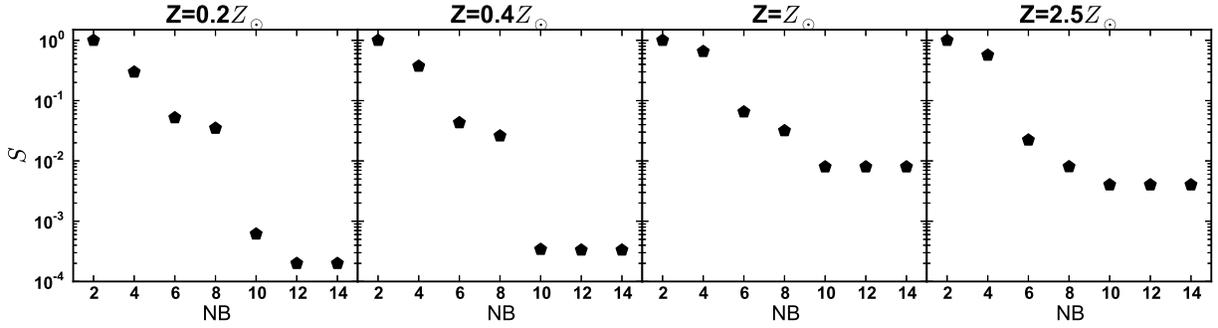}
\caption{Normalized median values of $S$ as a function of NB, the number of spectra in the basis.}
\label{ce_fig}
\end{figure}

In Figure \ref{ce_fig} we show the median of the $S$ distribution as a function of NB, the number of SSP spectra in the basis. We can see that the fits are improving (lower median) as we increase the number of elements in the basis until NB $\sim 10- 12$ where $S$ stabilises reaching a plateau, which indicates that adding more spectra to the basis will not significantly improve the ability to reproduce the remaining ones and will only add degeneracies to the problem being solved. Thus we select as an optimal basis the one with NB=12, the basis with the minimum number of spectra for which $S$ has reached this plateau for all model metallicities. 
 
The selected ages in our optimal 12-dimensional basis are: 
$1.78\times10^{5}$, 
$4.17\times10^{6}$, 
$8.71\times10^{6}$, 
$1.91\times10^{7}$, 
$3.60\times10^{7}$, 
$9.05\times10^{7}$, 
$3.60\times10^{8}$, 
$1.28\times10^{9}$, 
$2.75\times10^{9}$, 
$4.75\times10^{9}$,
$9.25\times10^{9}$ and
$13.75\times10^{9}$ yr.

A point worth noting is that a hand picked basis can lead to systematic biases in the recovered SFH.
We strongly encourage users of this sort of methods to try to find some criteria that allow them to 
reduce the degeneracy and minimize biases in the results.

\begin{figure}
\plotone{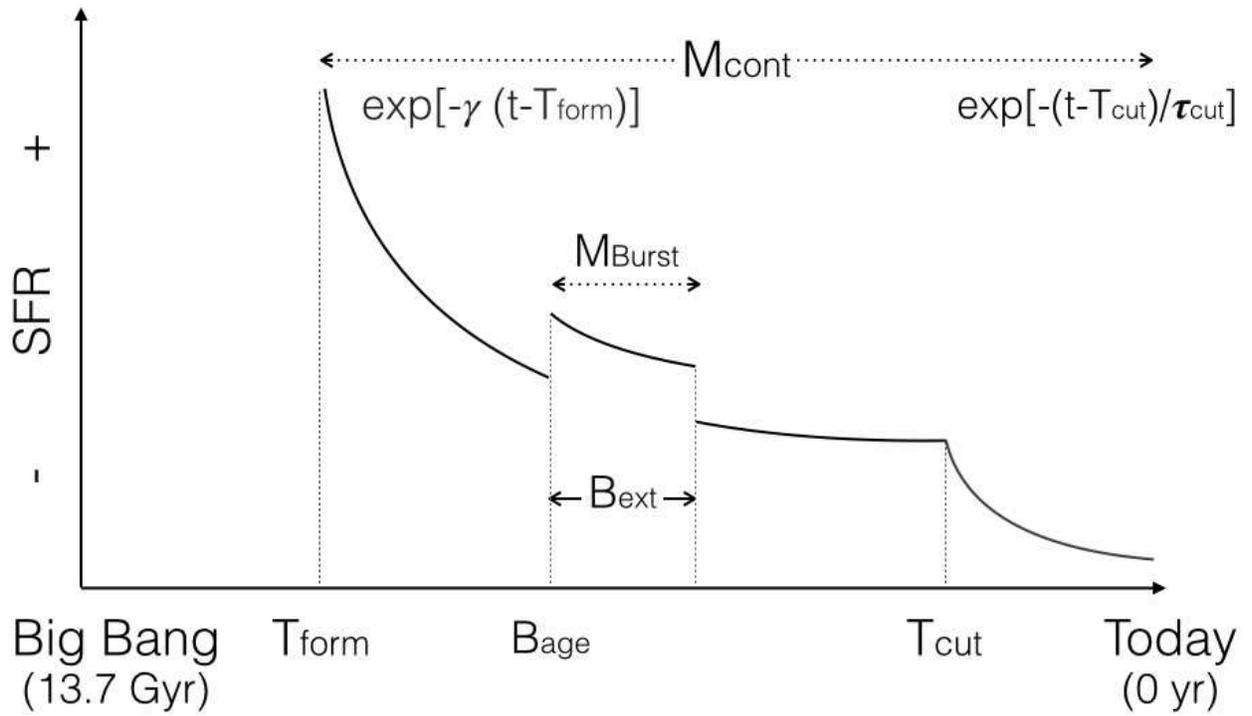}
\caption{
Sketch of a truncated SFR from \ssag, showing the burst and the truncation event.
}
\label{sketch}
\end{figure}

\begin{figure}
\plotone{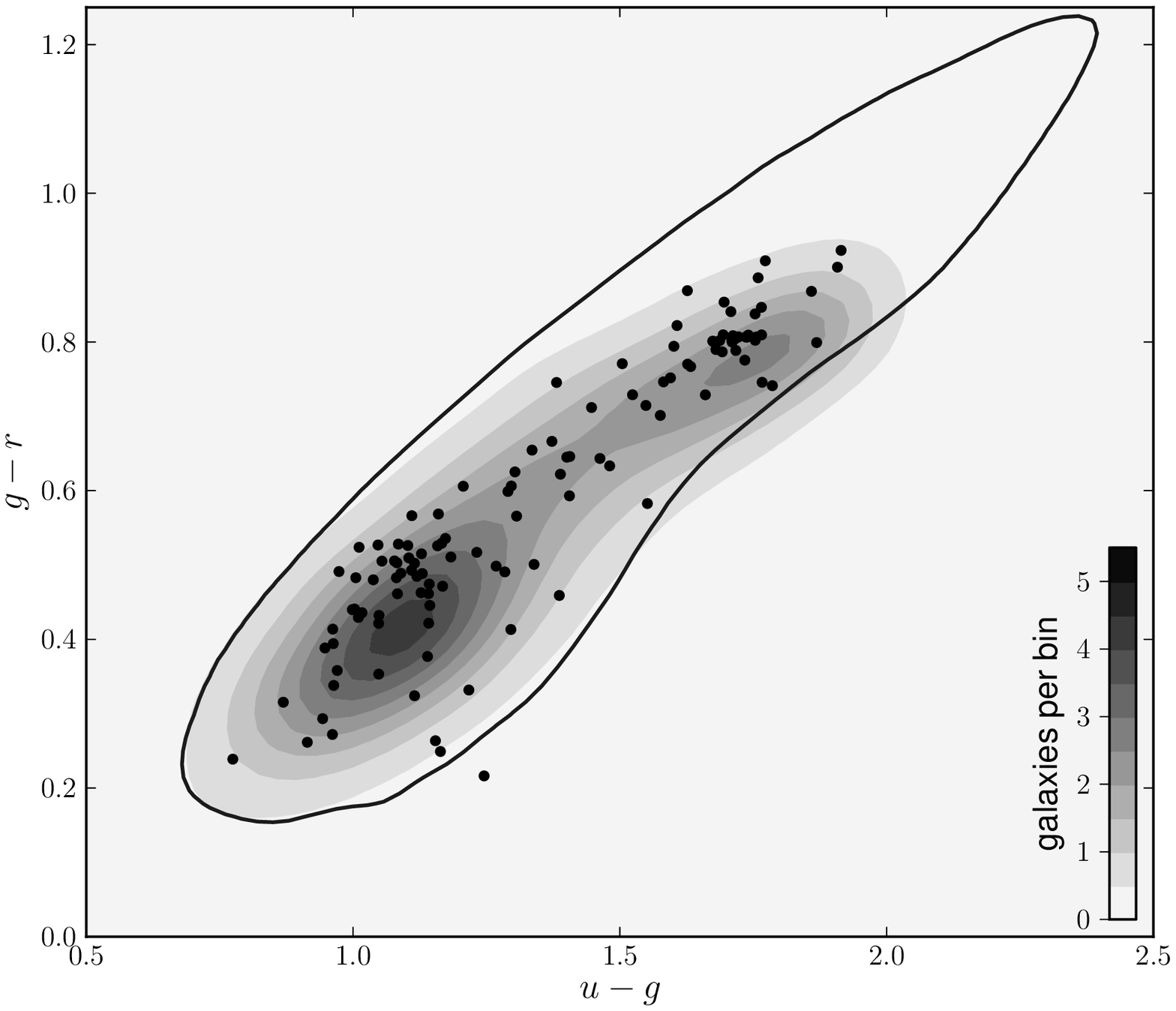}
\caption{The $(u-g,g-r)$ space density of SDSS-DR7 galaxies at $z<0.03$ is represented by the grey-shaded contours, in bins of (0.035, 0.021) mags and smoothed with a 2-$\sigma$ Gaussian kernel. The solid line encloses the total extent of \ssag~galaxies.
A subset of 120 \ssag~galaxies (black dots) was selected reproducing the bimodality shown by the SDSS galaxies.}
\label{sampled}
\end{figure}

\section{Synthetic Spectral Atlas of Galaxies (\ssag)}

As part of this investigation, we assembled an atlas of synthetic spectra of galaxies (\ssag)\footnote{Available at \url{http://www.astro.ljmu.ac.uk/~asticabr/SSAG.html}.}, containing a comprehensive set of 100,000 spectra.
These spectra were built from BC03 models using physically motivated SFHs that follow the prescription of \citet{chen12},  adapted as follows.
\begin{enumerate}

\item SFR:  In their recipe, \citet{chen12} assume that a galaxy forms stars according to an exponentially declining SFR and that at some time
in its life experiences a burst of star formation. A number of galaxies may undergo another event, 
after which the SFR starts to decrease at a faster rate than before.
We refer to this last event as  ``truncation". 
The following parameters, based on the  \citet{chen12} recipe, define the SFRs used to build SSAG (see Figure~\ref{sketch}).

\begin{itemize}

\item T$_{\mbox{form}}$: Age at which the galaxy starts forming stars. T$_{\mbox{form}}$ is uniformly distributed between the Big Bang (13.7 Gyr ago) and 1.5 Gyr.

\item $\gamma$: Characterises the decline of the exponential SFR ($\propto\exp^{-\gamma t}$). $\gamma$ is uniformly distributed between 0 and 1 Gyr$^{-1}$.

\item M$_{\mbox{cont}}$: Mass formed by the continuos SFR (i.e. without the burst contribution), obtained integrating the SFR from T$_{\mbox{form}}$ to the present.

\item B$_{\mbox{age}}$: Starting age of the star formation burst. The burst can start at any time in the past, with the constraint that 15 per cent of the galaxies experience
the burst in the last 2 Gyr.

\item B$_{\mbox{ext}}$: Duration of the burst episode, uniformly distributed in age between $3\times10^7$ and $3\times10^8$ yr. 
During B$_{\mbox{ext}}$ a constant SFR is superimposed on the underlying exponential SFR.
M$_{\mbox{burst}}$ is the stellar mass formed during the burst event (excluding the mass formed by the underlying exponential SFR).
The value of M$_{\mbox{burst}}$ is determined by the parameter A = M$_{\mbox{burst}}$/M$_{\mbox{cont}}$; log A is distributed uniformly between 0.03 and 4.

\item T$_{\mbox{cut}}$: Starting age of the truncation event. Following \citet{chen12}, we assume that 30 per cent of the galaxies experience a truncation event.
T$_{\mbox{cut}}$ is distributed in such a way that 10 per cent of the truncation events occur in the last 2 Gyr. 
After T$_{\mbox{cut}}$ the exponential SFR declines faster, with $e$-folding time $\tau_{\mbox{cut}}$; 
$\log \tau_{\mbox{cut}}$ is distributed uniformly between 7 and 9.

\end{itemize}

\item Metallicity: Each galaxy in \ssag~is built using a single metallicity. The metallicity is distributed uniformly in $Z/Z_\odot$ in such a way that 5 per cent of the galaxies
have $0.05 < Z/Z_\odot < 0.2$, and the remaining 95 per cent have $0.2 < Z/Z_\odot < 2.5$.

\item Dust extinction: The two-parameter \cite{cf00} prescription is used to model dust extinction in the \ssag~galaxies.
$\tau_{V}$, the optical depth in the $V$ band for stars younger than $10^7$ yr, takes values between 0 and 6.
$\mu$, the fraction of the optical depth resulting from the ISM, ranges from 0 to 2.
Inside these ranges the parameters follow Gaussian probability distribution functions of (mean, sigma) = (1.2, 0.985) and (0.3, 0.365), respectively.

\item Velocity dispersion: We convolve a Gaussian filter to the \ssag~spectra to simulate the effects of the stellar velocity dispersion, characterized by $\sigma_{v}$,
which take values distributed uniformly between 50 and 400 km~s$^{-1}$

\end{enumerate}
Once a set of input parameters is specified, we use the $csp\_galaxev$ code distributed by Bruzual \& Charlot to compute the
corresponding \ssag~spectrum. As input to this code we use the BC03 SSP models for the Padova 1994 stellar evolutionary tracks, 
the STELIB stellar atlas, and the \citet{chabrier03} IMF (see BC03 for details and references). 
When necessary, the BC03 models are interpolated in $\log Z$ to create an SSP model of the required $Z/Z_\odot$. Note that in \ssag~every SFH has a single value of metallicity contrary to \citet{chen12}. Also, for the synthesis of SSAG spectra we used SSP models corresponding to \citet{chabrier03} IMF instead of the \citet{kroupa01} IMF used in \citet{chen12}.

In Figure \ref{sampled} the solid line encloses the region occupied by all the \ssag~ galaxies in the $(u-g, g-r)$ color-color plane.
The shaded area shows the distribution of $z < 0.03$ SDSS DR7 galaxies \citep{abazajian09}.
The black dots in this Figure represent the 120 \ssag~galaxies used in this investigation.

The \ssag~will be made publicly available in electronic form after paper acceptance, together with a detailed description of the software used to build and read it, 
its file structure, and the physical parameter distributions.

%
%
%
%
%

\end{document}